\documentclass[11pt]{article}
\usepackage[a4paper,margin=1in]{geometry}
\usepackage{amsmath,amssymb,bm,mathtools}
\usepackage{graphicx}
\usepackage{booktabs}
\usepackage{enumitem}
\usepackage{authblk}
\usepackage[hidelinks]{hyperref}
\usepackage{physics}
\usepackage{setspace}
\usepackage{caption}
\usepackage{subcaption}
\usepackage[backend=biber,style=numeric,sorting=none,maxbibnames=99]{biblatex}
\usepackage{rotating}
\usepackage{array}
\usepackage{tabularx}
\usepackage{makecell}

\title{
A Synchronized Spin Model for Black-Hole Accretion Systems
}
\author[1]{Masahiro Morikawa}
\affil[1]{Riken, Ochanomizu University}
\author[2]{Akika Nakamichi}
\affil[2]{Kyoto-Sangyo-University}
\date{\today}

\begin{document}
\maketitle
\begin{abstract}
Black-hole accretion systems exhibit a characteristic coexistence of activities: broad-band
X-ray variability, hot coronae, wide-angle winds, and both steady and
discrete jets. This coexistence suggests a persistently time-dependent magnetic background in which
noisy fluctuations and explosive release are both essential. 
In this paper, we connect them all to intermittent magnetic reconnection and propose a Synchronized Spin Model (SSM) in which multiple local dynamos in a rotating accretion flow are represented as interacting macro-spins. 
Their synchronization, partial synchronization, excursion, and reversal define a compact set of collective variables that organize
both timing statistics and large-scale morphology. In this picture, multiscale magnetic reconnection sustains coronal heating, flares, intermittent outflows, and discrete jet activity, while
the same synchronization dynamics produce amplitude modulation and demodulation, providing a route to $1/f$-like variability, rms--flux/Taylor-like scaling, and approximately log-normal
statistics of the demodulated envelope. We further argue that, although the continuous flux
distribution in black-hole systems is more naturally discussed in multiplicative or log-normal
terms, broader event-catalog statistics remain useful for describing suitably defined burst hierarchies, particularly by analogy with solar and stellar flare systems. The hard/soft cycle of X-ray binaries is then interpreted as motion through magnetic state space.
\end{abstract}

\tableofcontents

\section{Introduction}

Black-hole accretion systems show a recurring combination of phenomena: broad-band X-ray variability, 
hot coronae, winds, compact jets, and transient ejecta. In X-ray binaries (XRBs), these observables reorganize across the hard, intermediate, and soft states; in active galactic nuclei (AGN), analogous spectral and outflow changes occur on much longer timescales, including in changing-look AGN \cite{FenderBelloniGallo2004,Belloni2010,Fender2010,LaMassa2015,MacLeod2016}. The main theoretical challenge is therefore not to explain any one component in isolation, but to identify a compact physical language that links timing, topology, and outflow morphology.

Existing approaches address important subsets of the problem. Propagating-fluctuation models reproduce red-noise continua and multiplicative variability \cite{Lyubarskii1997,Uttley2005,ArevaloUttley2006}.
Ordered-flux models explain how jets and winds are launched once a suitable magnetic geometry exists \cite{BlandfordZnajek1977,BlandfordPayne1982}. Reconnection and plasmoid pictures explain impulsive dissipation, flares, and intermittent ejections \cite{Aschwanden2011SOC,GoodmanUzdensky2008,ComissoSironi2022}. What remains less clear is why these observables reorganize together.

We address this problem with a \emph{Synchronized Spin Model} (SSM). The model is not intended as a microscopic alternative to General Relativistic Magnetohydrodynamics (GRMHD). Instead, it is a mesoscopic effective description in which the inner accretion flow is coarse-grained into interacting magnetic domains, or ``macro-spins.'' Their alignment, partial synchronization, excursion, and reversal define a small set of collective variables that organize both the timing statistics and the magnetic topology of the flow.

The paper proceeds as follows. 
Section~2 summarizes the observational constraints. 
They motivate us to introduce the coarse-grained spin variables and the Synchronized Spin Model (SSM) in Section~3. Then we discuss the statistical consequences of the spin synchronization in Section~4. 
SSM maps the magnetic-state variables to coronae, flares, winds, and jets, as shown in Section~5. 
Section~6 describes how the SSM reproduces the observed statistics. 
Section~7 connects the model to XRB cycles and the q-diagram. 
Section~8 describes observational falsifiability of the SSM
scenario. 
Section~9 discusses the SSM's niche within existing frameworks, its limitations, and possible validation routes.
The final section~10 concludes this work. 

\section{Observational Constraints from Black-Hole Systems}

A unified magnetic-state model must account for at least six empirical regularities.

\subsection*{(i) Broad-band variability} 
\begin{figure}
    \centering
    \includegraphics[width=1\linewidth]{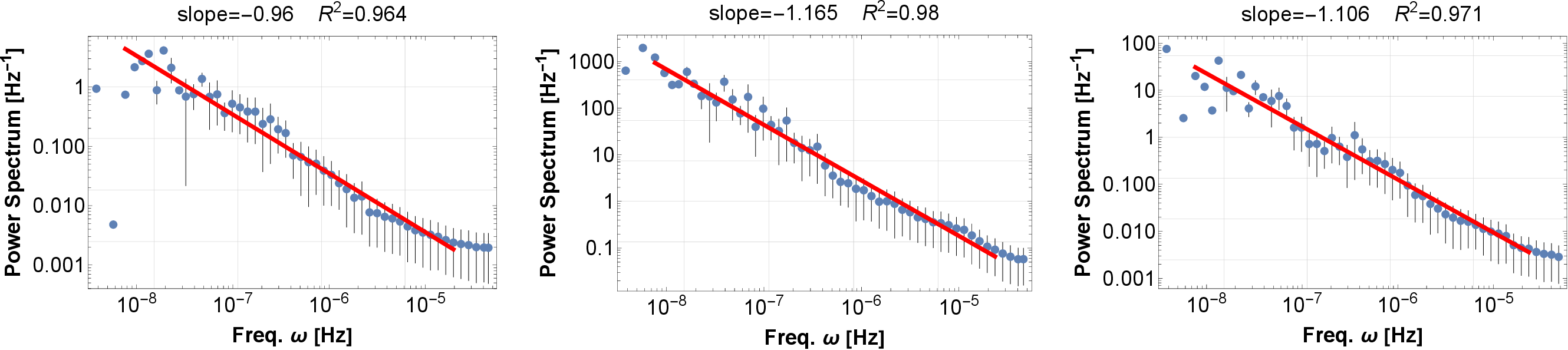}
    \caption{Representative power spectral densities (PSD) derived from 16 yr MAXI X-ray light curves for an AGN (Mrk 421; left), a black-hole X-ray binary (Cyg X-1; center), and a neutron-star X-ray binary (1A 1742-294; right). Approximately one-third of the several hundred MAXI sources examined show pink broad-band variability across part of the band. 
    \newline
    This figure is obtained as follows. The original MAXI X-ray flux data are linearly interpolated to produce a time series with equal time intervals. After PSD is calculated using DFT, the data are averaged within each octave, with error bars. Then we obtain the sliding-fit window that maximizes the Coefficient of Determination \(R^2\). The procedure is the same for all the PSD graphs in this paper.   
    }
    \label{fig:1}
\end{figure}

Many accreting black-hole systems show power spectral densities (PSDs) of the form
\begin{equation}
P(f)\propto f^{-\alpha},
\qquad \alpha\sim 1,
\label{eq:psdobs}
\end{equation}
where $P(f)$ is the Fourier power at temporal frequency $f$ and $\alpha$ is the effective low-frequency slope. This behavior persists across part of the band, although the slope is state-dependent and not universal \cite{Lyubarskii1997,Uttley2005,HeilVaughan2010}. Figure~\ref{fig:1} illustrates representative MAXI power spectral density (PSD), 
for an AGN, a black-hole XRB, and a neutron-star XRB. The presence of 1/f variability in both black-hole and neutron-star accretors argues that the long-memory component is primarily a property of the accretion disk and not of the central compact objects.       

\subsection*{(ii) Continuous-flux statistics and event-catalogue statistics}

We distinguish two related but non-identical statistical levels.

First, for the continuous X-ray light curve, many accreting black-hole systems are more
naturally characterized by multiplicative variability, a linear rms--flux relation, and an
approximately log-normal flux distribution than by a single scale-free law for the instantaneous
flux itself \cite{Uttley2005,Gleissner2004CygX1,HeilVaughan2010}. A useful
phenomenological representation is therefore
\begin{equation}
 p(F)=\frac{1}{F\,\sigma_{\ln F}\sqrt{2\pi}}
 \exp\!\left[-\frac{(\ln F-\mu_{\ln F})^2}{2\sigma_{\ln F}^2}\right],
\label{eq:lognormal_flux}
\end{equation}
where $F$ is the flux and $\mu_{\ln F}$ and $\sigma_{\ln F}$ are the mean and standard deviation
of $\ln F$. This form is the observationally better-motivated default for the continuous
light-curve statistics of black-hole accretion systems.

Second, if one decomposes the variability into threshold-defined shots, flares, or ejecta, the resulting event catalog can, in some systems or analyses, exhibit broad, approximately
scale-free energy or amplitude statistics,
\begin{equation}
 N(E) \propto E^{-\alpha_E},
\label{eq:event_powerlaw}
\end{equation}
where $N(E)$ is the occurrence rate of events with energy or amplitude proxy $E$ and
$\alpha_E$ is the corresponding index. Such power-law-like event statistics are particularly
common in solar and stellar flare catalogues \cite{Maehara2015,Aschwanden2011SOC},
whereas in black-hole systems, the inferred event distribution depends more strongly on how events are defined and should not be conflated with the flux distribution of the continuous
light curve \cite{Uttley2005
}.


\subsection*{(iii) Steady and transient jets} XRBs show both compact steady jets and discrete ejecta, especially near the hard-to-soft transition \cite{FenderBelloniGallo2004,Fender2010}. This motivates a decomposition into a quasi-steady and a transient channel,
\begin{equation}
P_{\rm jet}^{\rm steady}, \qquad P_{\rm jet}^{\rm disc},
\label{eq:jetpair}
\end{equation}
where the superscripts denote steady compact-jet power and discrete ejection power. The two channels need not respond identically to magnetic order or state change.

\subsection*{(iv) Persistent coronae} Coronal emission is long-lived yet highly variable. A useful coarse-grained description is
\begin{equation}
L_{\rm cor}\sim \int_V Q_{\rm diss}\,dV,
\qquad
Q_{\rm diss}\sim \eta_{\rm eff}J^2+Q_{\rm turb}+Q_{\rm rec},
\label{eq:lcorobs}
\end{equation}
where $L_{\rm cor}$ is the coronal luminosity, $V$ is the emitting volume, $Q_{\rm diss}$ is the local dissipation rate, $J$ is the current density, $\eta_{\rm eff}$ is an effective resistivity, and $Q_{\rm turb}$ and $Q_{\rm rec}$ denote unresolved turbulent and reconnection heating. The point of Eq.~(\ref{eq:lcorobs}) is not precision modeling, but to emphasize that the corona can be sustained by many small dissipative sites rather than by a single coherent event \cite{Fabian2015,GoodmanUzdensky2008}.

\subsection*{(v) Structured winds.} AGN and disk winds are often modeled as continuous outflows, but observations increasingly show clumpy or intermittent structure \cite{KingPounds2015,Tombesi2010,Fukumura2017}. It is therefore useful to separate a continuous background from a discrete magnetic release,
\begin{equation}
P_{\rm wind}=P_{\rm wind}^{\rm cont}+P_{\rm wind}^{\rm disc},
\label{eq:windpair}
\end{equation}
where $P_{\rm wind}^{\rm cont}$ and $P_{\rm wind}^{\rm disc}$ denote continuous and discrete wind power, respectively.

\subsection*{(vi) State transitions and excursions.} In XRBs, hardness, timing properties, jet behavior, and broad outflow signatures reorganize together along the HID/q-diagram \cite{Belloni2010}. In AGN, changing-look events indicate that large spectral changes need not be accompanied by equally strong new jets \cite{LaMassa2015,MacLeod2016}. Any successful model must therefore distinguish full large-scale magnetic reorganization from more limited excursions in magnetic state space.

These constraints motivate a nonequilibrium magnetic description in which timing statistics and outflow morphology arise from the same evolving collective variables.

\section{Physical Basis of the Synchronized Spin Model (SSM)}

We now describe our simple model, starting with the basic features common to all astronomical accretion-disk systems. 

\subsection{From rotating MHD to macro-spin}

\begin{figure}
    \centering
    \includegraphics[width=0.5\linewidth]{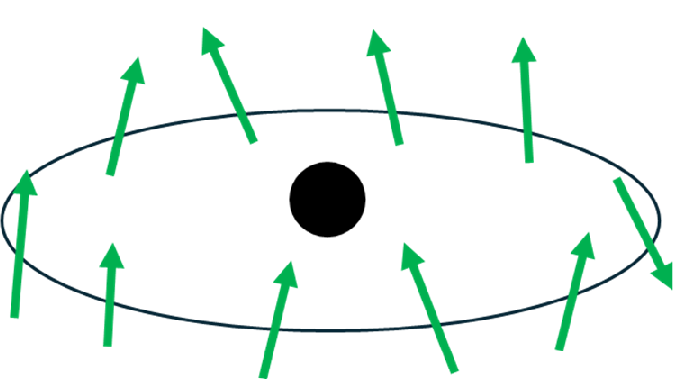}
    \caption{Schematic configuration of the SSM local dynamo domains (green) embedded in the inner part of the accretion disk. Each spin represents a coarse-grained magnetic domain or the winding current around a locally coherent column-like structure. Their collective alignment, frustration, and reversal determine the large-scale magnetic topology and its observable consequences.}
    \label{fig:2}
\end{figure}

The starting point is a rotating, conducting, magnetized flow described at the continuum level by resistive MHD,
\begin{equation}
\rho\left(\frac{\partial \mathbf{u}}{\partial t}+\mathbf{u}\cdot\nabla \mathbf{u}\right)
=-\nabla p+\rho\mathbf{g}+\frac{1}{4\pi}(\nabla\times\mathbf{B})\times\mathbf{B}+\nabla\cdot\mathbf{\Pi}-2\rho\,\mathbf{\Omega}\times\mathbf{u},
\label{eq:mom}
\end{equation}
with $\nabla\cdot\mathbf{B}=0$. Here $\rho$ is mass density, $\mathbf{u}$ is velocity, $p$ is pressure, $\mathbf{g}$ is gravitational acceleration, $\mathbf{B}$ is magnetic field, $\mathbf{\Pi}$ is viscous stress, $\mathbf{\Omega}$ is the rotation vector.
In a rotationally constrained regime,
\begin{equation}
Ro=\frac{U}{2\Omega L}\ll 1,
\label{eq:rossby}
\end{equation}
where $U$ and $L$ are representative velocity and length scales. 
When $Ro\ll 1$, the dominant Coriolis force (the last term in Eq.(\ref{eq:mom})) is balanced by the pressure gradient (the first term): \(-\nabla p=2\rho\+\mathbf{\Omega}\times\mathbf{u}\). 
Taking the curl of this balance yields 
\begin{equation}
\mathbf{\Omega}\cdot\nabla \mathbf{u}=0
\label{eq:taylorpre}
\end{equation}
i.e., the rotation can maintain columnar structures aligned with the rotational axis even in a turbulent flow. 
This motivates a Taylor-column-like picture of locally coherent dynamo domains, used here only as a coarse-grained prototype for the inner disk.
For each coarse-grained domain $V_i$, we define a magnetic moment-like variable
\begin{equation}
\mathbf{S}_i\equiv \frac{1}{V_i}\int_{V_i}\mathbf{r}\times\mathbf{J}\,dV,
\qquad \mathbf{J}=\frac{c}{4\pi}\nabla\times\mathbf{B},
\label{eq:Si}
\end{equation}
where $\mathbf{r}$ is the position vector measured from the domain center, $\mathbf{J}$ is current density, and $c$ is the speed of light. The quantity $\mathbf{S}_i$ is not the literal spin of a particle. It is a coarse-grained magnetic moment that measures the net handedness of current circulation within domain $V_i$. We then normalize it to define a macro-spin,
\begin{equation}
\mathbf{s}_i=\frac{\mathbf{S}_i}{|\mathbf{S}_i|},
\qquad |\mathbf{s}_i|=1.
\label{eq:si}
\end{equation}
The ensemble $\{\mathbf{s}_i\}$ is the basic mesoscopic state variable of the SSM. Figure~\ref{fig:2} shows the intended geometry: the spins occupy the inner part of the disk and represent coarse-grained magnetic dynamo domains rather than individual field lines.

\subsection{Effective interactions and collective dynamics}

Neighboring domains interact through induction, flux constraints, and magnetic stress transmission. A minimal Lagrangian may be written for the macro-spin variables $\{\mathbf{s}_i\}$, almost uniquely by demanding the rotational invariance,  
as \cite{Nakamichi2012MNRAS4232977}
\begin{align}
L &= K-V, \\
K &= \frac{1}{2}\sum_{i=1}^{N}\dot{\mathbf{s}}_i^{\,2}, \\
V &= \mu\sum_{i=1}^{N}(\mathbf{\Omega}\cdot\mathbf{s}_i)^2+\frac{\lambda}{2N}\sum_{i<j}\mathbf{s}_i\cdot\mathbf{s}_j,
\label{eq:Vspin}
\end{align}
where $N$ is the number of macro-spins, $K$ is an inertial kinetic term for slow collective motion, and $V$ is the effective interaction potential. The coefficient $\mu$ weights alignment with the rotation axis, while $\lambda$ measures the mean spin--spin coupling. Equation~(\ref{eq:Vspin}) is not derived from first-principles GRMHD. It is the lowest-order rotationally symmetric closure that favors either axis alignment or collective ordering and therefore serves as a reduced state model.

This SSM is a conservative system, in contrast to the real, highly dissipative system. Despite being conservative, SSM is fully chaotic and highly random, capturing some important aspects of the irregular collective behavior seen in dissipative systems.  
In fact, the dissipative version of SSM \cite{Mori2013PRE87012108} yields qualitatively similar magnetic activity.     

To see the origin of the representative evolution equation, write each spin in polar form relative to $\mathbf{\Omega}$ and consider the angular degree of freedom $\theta_i$. Variation of the Lagrangian then gives, schematically,
\begin{equation}
\ddot{\theta}_i
=
- \mu \Omega^2 \sin 2\theta_i 
- \frac{\lambda}{N} \sum_j \sin(\theta_i - \theta_j).
\label{eq:kuramoto}
\end{equation}
The first term comes from the derivative of $(\mathbf{\Omega}\cdot\mathbf{s}_i)^2\propto \cos^2\theta_i$ and therefore restores spins toward the rotation axis; either parallel or anti-parallel. The second term is the usual phase-coupling term obtained by differentiating pairwise dot products $\mathbf{s}_i\cdot\mathbf{s}_j=\cos(\theta_i-\theta_j)$. Equation~(\ref{eq:kuramoto}) is therefore a second-order synchronization equation closely related to the inertial Kuramoto model \cite{Ji2014,RajwaniJalan2025}. In the present paper, it is used only as a reduced synchronization language, not as a microscopic derivation from MHD.

\subsection{Order variables}

For synchronization theory, the physically useful morphological variable is the vector magnetization,
\begin{equation}
\mathbf{M}=\frac{1}{N}\sum_{i=1}^{N}\mathbf{s}_i,
\qquad M=|\mathbf{M}|.
\label{eq:Mvec}
\end{equation}
Large $M$ corresponds to a comparatively ordered, nearly dipolar state; small $M$ corresponds to a disordered or multipolar state.

To characterize reversal-prone boundaries, we define the reversal-gradient activity
\begin{equation}
\mathcal{G}=\int_V |\nabla \mathbf{M}(\mathbf{x},t)|^2\,dV,
\label{eq:Gdef}
\end{equation}
which is large when neighboring regions carry strongly different local magnetizations. To distinguish closed from open topology, we define the open-flux fraction
\begin{equation}
f_{\rm open}=\frac{\Phi_{\rm open}}{\Phi_{\rm tot}},
\label{eq:fopen}
\end{equation}
where $\Phi_{\rm open}$ is the magnetic flux threading field lines that escape the local system and $\Phi_{\rm tot}$ is the total unsigned flux. A fourth useful quantity is the axial alignment factor
\begin{equation}
C_{\rm ax}=\frac{\mathbf{M}\cdot\mathbf{\Omega}}{|\mathbf{M}|\,|\mathbf{\Omega}|},
\label{eq:Cax}
\end{equation}
which measures how well the ordered component aligns with the rotation axis. The set $(M,\mathcal{G},f_{\rm open},C_{\rm ax})$ will organize the phenomenology below.

\begin{figure}
    \centering
    \includegraphics[width=1\linewidth]{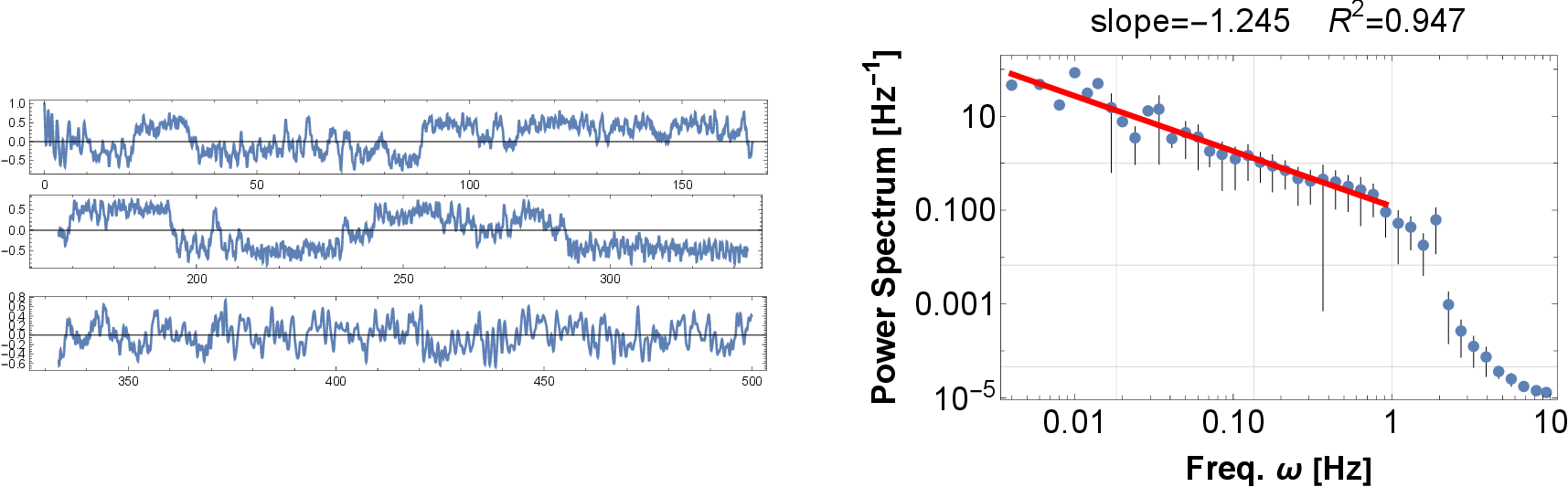}
    \caption{
    A typical SSM calculation result. 
    Parameters are: \(\mu=-10, \lambda=-20, N=9, K=-V/2, \mathrm{time}=500 \). 
    Left: time series of the projection of the mean spin onto the rotation axis, $C_{ax}$. The three stacked panels represent successive time segments of a single continuous run. Long intervals of one polarity are interrupted by rapid reversals; near-failures of reversal correspond to excursion-like events. Right: power spectral density (PSD) of the same series.
    }
    \label{fig:3}
\end{figure}

\begin{figure}
    \centering
    \includegraphics[width=0.75\linewidth]{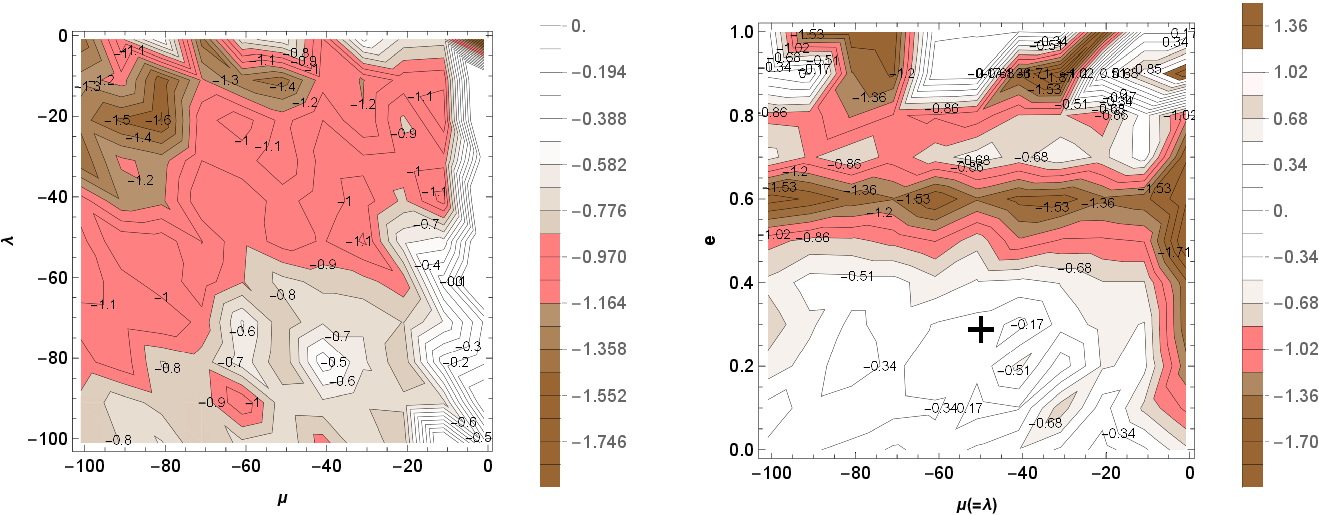}
    \caption{Distribution of the PSD power-law index in SSM parameter space. 
    Left: map in the $(\mu,\lambda)$ plane, with \(K=-V/2\), where the pink-noise regime occupies a broad region, indicating that the pink noise is not a critical phenomenon. 
    Right: map in the plane of $\mu(=\lambda)$ and the initial energy in the form of $e=1+(K/V)$. All the initial spins are almost up, and the potential energy is close to its minimum. Thus, $e=1$ means almost no initial kinetic energy, and $e=0$ means mild initial kinetic energy. The pink region becomes narrower in this representation, motivating the more detailed analysis in Fig.~\ref{fig:5}, where the point marked "+" is examined.}
    \label{fig:4}
\end{figure}

\begin{figure}
    \centering
    \includegraphics[width=1\linewidth]{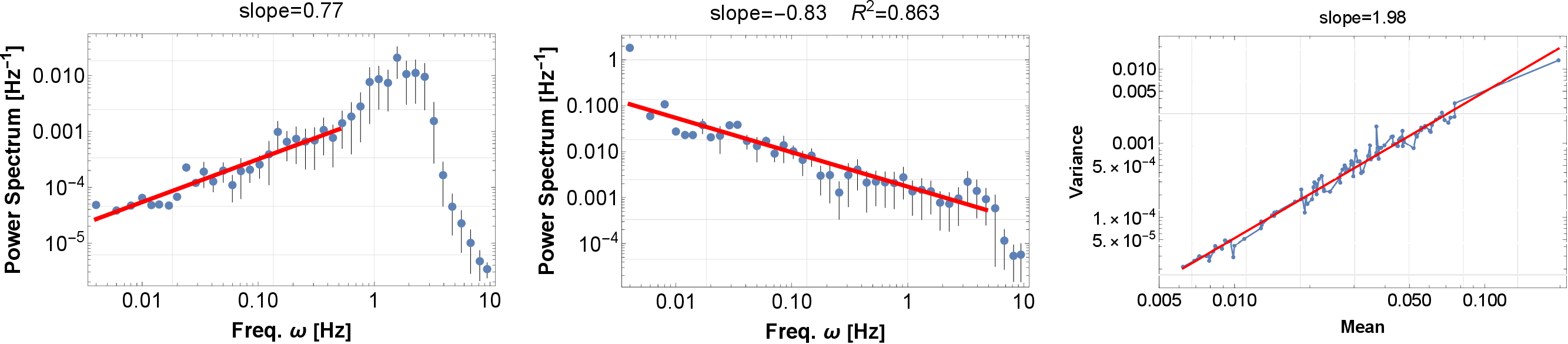}
    \caption{Statistical diagnostics for a representative parameter choice (\(\mu=-50, \lambda=-50, N=10, e=0.3, \mathrm{time}=500 \) (the point marked "+" in Fig. \ref{fig:4})) from the pale/white-noise-like region in the right panel of Fig.~\ref{fig:4}. 
    Left: PSD of the detrended SSM data, which is not pink. Middle: PSD of the absolute value of the detrended data, which becomes pink, consistent with a demodulated amplitude envelope. 
    Right: Taylor’s law test. The raw and merely detrended data do not follow Taylor's law, whereas the absolute value of the detrended series does.}
    \label{fig:5}
\end{figure}

\begin{figure}
    \centering
    \includegraphics[width=0.8\linewidth]{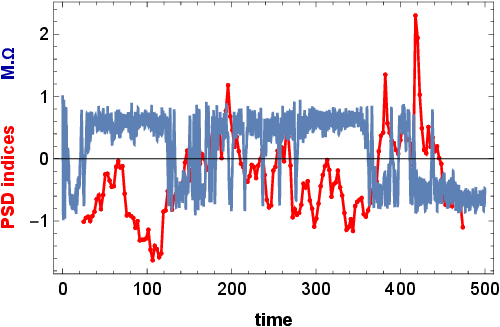}
    \caption{
    The spin magnetic parameter $C_{ax}$ (blue) and the running PSD indices (red) for the SSM calculation with the same parameter as for Fig.\ref{fig:3}, but another run. 
    The running PSD indices are computed in overlapping windows whose length is 10\% of the full record and whose start times are shifted by 1\% each step. 
    The PSD index tends to increase during excursions or polarity reversals, suggesting that global desynchronization weakens frequency crowding and therefore suppresses pink-noise behavior.}
    \label{fig:6}
\end{figure}

The numerical behavior of the model is illustrated in Figs.~\ref{fig:3}--\ref{fig:6}. 
We now explain them very briefly. The details will be described in the following sections. 
Figure~\ref{fig:3} shows the time series of the projection of the mean spin onto the rotation axis and its PSD. Long polarity epochs are interrupted by rapid reversals, while smaller excursion-like events occur without a full sign change. Figure~\ref{fig:4} maps the PSD slope over parameter space and shows that pink-noise-like behavior occupies a broad region. Figure~\ref{fig:5} demonstrates that for some parameter choices the raw detrended series is not pink, whereas the absolute value of the detrended series becomes pink and satisfies Taylor's law, consistent with an amplitude-modulation/demodulation interpretation. Figure~\ref{fig:6} shows that the local PSD index tends to increase during polarity reversals, suggesting that strong global disorder suppresses frequency crowding at those times.

\subsection{Reversal, excursion, and reconnection}
In a finite-$N$ SSM, the polarity branches $+M$ and $-M$ should be regarded as metastable. Since the full system is conservative but strongly chaotic, the unresolved internal degrees of freedom act as an effective finite-size bath for the collective magnetization. As a result, polarity reversals can happen as a result of chaos-assisted transitions between two metastable branches. In this sense, polarity reversal in SSM is a natural consequence of finite-$N$ nonlinear collective dynamics. Only in the formal $N\to\infty$ limit would one expect a truly frozen spontaneous-symmetry-broken branch.

The model distinguishes a full global reversal,
\begin{equation}
\mathbf{M}\rightarrow -\mathbf{M},
\label{eq:reversal}
\end{equation}
from a large excursion in which $M$ is strongly disturbed but does not settle into the opposite polarity. This distinction will later be important for comparing XRB transitions with changing-look AGN.

Magnetic reconnection enters the model as a multiscale energy-conversion channel. A dimensional estimate for the released power is
\begin{equation}
P_{\rm MR}\sim \int_{A_{\rm cs}} \frac{B^2}{4\pi}v_{\rm in}\,dA,
\qquad
v_{\rm in}=\epsilon_{\rm rec}v_A,
\label{eq:PMR}
\end{equation}
where $A_{\rm cs}$ is the total reconnecting current-sheet area, $B^2/4\pi$ is the magnetic energy density, $v_{\rm in}$ is the inflow speed into the sheet, $\epsilon_{\rm rec}$ is the dimensionless reconnection rate, and $v_A$ is the local Alfv\'en speed. 
Here $v_{\rm in}$ is normalized by the local Alfv\'en speed, not because the inflow itself is an Alfv\'en wave, but because the reconnection exhaust is accelerated by magnetic tension to a speed of order $v_A$, and the inflow speed is a fraction of that characteristic MHD response speed.
The factor $B^2v_{\rm in}/4\pi$ is the incoming Poynting-flux scale into the sheet, so Eq.~(\ref{eq:PMR}) estimates how rapidly magnetic energy can be converted into plasma heating, bulk acceleration, and non-thermal particles \cite{GoodmanUzdensky2008,ComissoSironi2022}.

\section{Statistical Structure of the SSM}

\subsection{From synchronization to 1/f-like variability}

The central statistical idea is that dynamic synchronization produces long envelopes through frequency crowding and beating. In previous work, we argued that such envelope formation, followed by nonlinear demodulation, naturally produces pink-noise-like behavior \cite{MorikawaNakamichi2023}. The present SSM exhibits exactly that structure, which is why synchronization is central to the model.

Two nearby frequencies generate a beat envelope when 
$\omega\gg\lambda$:
\begin{equation}
\cos[(\omega+\lambda)t]+\cos[(\omega-\lambda)t]=2\cos(\lambda t)\cos(\omega t),
\label{eq:AM}
\end{equation}
where $\omega$ is the fast carrier frequency and $\lambda$ is the beat frequency. Thus, a synchronized ensemble with many small frequency differences automatically produces low-frequency modulation.

If the collective signal is written as
\begin{equation}
X(t)=\sum_k A_k\cos(\omega_k t+\phi_k),
\label{eq:Xsum}
\end{equation}
with amplitudes $A_k$, phases $\phi_k$, and clustered frequencies $\omega_k$, then partial synchronization generates a hierarchy of envelope timescales. When the hierarchy of cluster sizes is broad, the envelopes naturally populate many timescales and produce pink continua rather than a single preferred period. 
In particular, if the spacing of the dominant modulation scales is approximately exponential, the demodulated envelope approaches a $1/f$ spectrum \cite{MorikawaNakamichi2023}. This is why the absolute value or square of the detrended signal is the relevant quantity in Fig.~\ref{fig:5} and later in Fig.~\ref{fig:9}.

This amplitude-modulation mechanism for 1/f-like variability does not require that all of the observed low-frequency structure be attributed to a pre-existing hierarchy of intrinsic timescales. On the other hand, most existing explanations rely on a broad hierarchy of intrinsic timescales across the observed band \cite{Lyubarskii1997,Uttley2005,ArevaloUttley2006}.

\subsection{Taylor's law and the rms--flux relation}
\label{sec:taylor-ssm}
A scaling analogous to Taylor’s law follows naturally from the amplitude-modulation picture. In Eq.~(\ref{eq:AM}), the high-frequency carrier $\cos(\omega t)$ is modulated by the slowly varying factor $\cos(\lambda t)$. 
In the same spirit, we model the short-time increment of any observed quantity $x_t$ as
\begin{equation}
dx_t=A_t\,d\xi_t,
\label{eq:dxAM}
\end{equation}
where $A_t$ is a slowly varying envelope signal and $d\xi_t$ is a faster carrier or noise component without any structure. Within a time window $T$ short enough that $A_t$ is approximately constant, the local variance becomes
\begin{equation}
\mathrm{Var}_T(dx)
\sim
\left\langle (dx)^2 \right\rangle_T
=
A_t^2 \left\langle (d\xi_t)^2 \right\rangle_T,
\label{eq:varlocal}
\end{equation}
while the local mean amplitude of the demodulated signal scales as
\begin{equation}
\mu_T
\sim
\left\langle |dx| \right\rangle_T
=
A_t \left\langle |d\xi_t| \right\rangle_T.
\label{eq:mulocal}
\end{equation}
If the short-time statistics of $d\xi_t$ are approximately stationary within the window, then both $\langle (d\xi_t)^2\rangle_T$ and $\langle |d\xi_t|\rangle_T$ are effectively constant there, so eliminating $A_t$ between Eqs.~(\ref{eq:varlocal}) and (\ref{eq:mulocal}) yields
\begin{equation}
\mathrm{Var}_T(dx) \propto \mu_T^2.
\label{eq:taylorlaw}
\end{equation}
This gives a natural Taylor-like fluctuation scaling for the demodulated series. Here, the ``mean'' quantity is the local mean amplitude of the demodulated signal rather than the signed mean increment itself.
When this variable $x$ represents the energy flux $F$, this makes the connection to Taylor-like scaling explicit in astrophysical variability studies.

In black-hole accretion variability, the same phenomenon is usually discussed as the rms--flux relation rather than under the name of Taylor's law \cite{Uttley2005,HeilVaughan2010}: 
\begin{equation}
\sigma_{\rm rms} = k\,( \bar{F} - C ), 
\label{eq:rms-flux}
\end{equation}
where \(\bar{F}\) is the average flux over that time interval, \(\sigma_{\rm rms}\) is the absolute root-mean-square (rms) variability amplitude within that interval, \(C\) is the trend, \(k\) is a constant. 
If we express this in terms of variance (\(=\sigma_{\rm rms}^2\)), this is the same form as Taylor's law Eq.(\ref{eq:taylorlaw}). 
The present point is that the demodulated envelope can show the same quadratic scaling without separately postulating a wide deterministic hierarchy of low frequencies. The demodulation behavior seen in Fig.~\ref{fig:5} supports this viewpoint.

\subsection{From amplitude modulation/demodulation to log-normal variability and power-law tails}\label{sec:lognormal-ssm}

Section~\ref{sec:taylor-ssm} presented a minimal phenomenological route from amplitude modulation/demodulation to multiplicative variability. Here we recast the same idea in a form that is more consistent with the conservative SSM, and that also prepares the reduced stochastic description used again in Sec.~8.

The full SSM is conservative, so its total Hamiltonian is constant. However, after separating a slow collective phase from fast internal degrees of freedom,
\begin{equation}
\theta_i(t)=\Phi(t)+\delta_i(t),
\end{equation}
the collective sector can exchange energy with the chaotic internal sector $\{\delta_i\}$. Therefore, the relevant stochastic variable is not the conserved total energy of the full SSM, but the partial energy assigned to the slow collective mode,
\begin{equation}
E\equiv E_{\Phi}=\frac{1}{2}\dot{\Phi}^{2}+\frac{1}{2}\omega_{0}^{2}\Phi^{2}.
\label{eq:Ephi_def_rewrite}
\end{equation}
Near a locally stable branch, coarse-graining of the fast chaotic variables yields an effective second-order Langevin equation for the slow mode,
\begin{equation}
\ddot{\Phi}+\bigl[\omega_{0}^{2}+\eta(t)\bigr]\Phi=\xi(t),
\label{eq:Phi_effective_rewrite}
\end{equation}
where $\eta(t)$ and $\xi(t)$ are not external noises but effective fluctuations generated by the internal SSM dynamics. In the notation introduced above,
\begin{equation}
\eta(t)=2\mu\Omega^{2}\bigl[C_{2}(t)-\overline{C}_{2}\bigr],
\qquad
\xi(t)=-\mu\Omega^{2}S_{2}(t)-\overline{\ddot\delta}(t),
\label{eq:eta_xi_compact}
\end{equation}
with
\begin{equation}
C_{2}(t)\equiv \frac{1}{N}\sum_{i=1}^{N}\cos 2\delta_i,
\qquad
S_{2}(t)\equiv \frac{1}{N}\sum_{i=1}^{N}\sin 2\delta_i.
\end{equation}
In particular, $\eta(t)$ represents a multiplicative random force, 
while $\xi(t)$ describes the standard additive random force. 

Instead of detailing the transformation from Eq.~\eqref{eq:Phi_effective_rewrite} to an effective energy equation, we only summarize the result. Following the standard energy--angle reduction for noisy oscillators, one rewrites $(\Phi,\dot\Phi)$ in terms of $(E,\chi)$, averages over the fast angle $\chi$, and then coarse-grains over times longer than the bath correlation time. This yields an effective first-order stochastic dynamics for $E$,
\begin{equation}
\dot{E}=aE+bE\eta+c\sqrt{E}\xi.
\label{eq:E_effective_general}
\end{equation}
The multiplicative term \(bE\eta\) controls the finite-time log-normal regime, whereas the additive reinjection term \(c\sqrt{E}\xi\) regularizes the process and allows stationary heavy tails.
In the multiplicative-dominated regime, the last term drops in Eq.~\eqref{eq:E_effective_general}. Applying Ito's lemma to $\ln E$ gives  
\begin{equation}
\frac{d{\ln E}}{dt}=(a-b^2/2)+b\eta
\end{equation}
Hence $\ln E$ is Gaussian and the finite-time distribution of $E$ is log-normal,
\begin{equation}
P(E,t)=\frac{1}{E\sqrt{2\pi b^{2}t}}
\exp\!\left[
-\frac{\bigl(\ln E-\ln E_{0}-(a-b^{2}/2)t\bigr)^{2}}{2b^{2}t}
\right].
\label{eq:E_lognormal_rewrite}
\end{equation}
This is the conservative-SSM counterpart of the simpler amplitude-modulation argument of Sec.~4.3: demodulation isolates the slowly varying collective envelope, while the chaotic internal sector acts as an effective multiplicative bath.

For the stationary statistics, one must retain the reinjection term \(c\sqrt{E}\xi\). 
A simple one-variable Fokker--Planck description consistent with Eq.~\eqref{eq:E_effective_general} is
\begin{equation}
\frac{\partial P}{\partial t}
=-\frac{\partial}{\partial E}\bigl[aEP\bigr]
+\frac{1}{2}\frac{\partial^{2}}{\partial E^{2}}
\Bigl[\bigl(b^{2}E^{2}+c^{2}E\bigr)P\Bigr].
\label{eq:fp_E_rewrite}
\end{equation}
The large-$E$ tail is controlled by the multiplicative diffusion term $b^{2}E^{2}$. 
If the reduced one-variable process admits a stationary state with vanishing probability current, then the large-$E$ tail becomes algebraic,
\begin{equation}
P_{\rm st}(E)\propto E^{-\alpha},
\qquad
\alpha=2-\frac{2a}{b^{2}}.
\label{eq:E_powerlaw_rewrite}
\end{equation}
Thus, the same reduced description yields two complementary regimes: finite-time log-normal variability when multiplicative growth dominates, and a power-law stationary tail once reinjection is retained, and a stationary Fokker--Planck balance is imposed. This reduced stochastic picture will be used again in Sec.~8 as the effective one-variable description of the collective mode extracted from the conservative many-body SSM.


\subsection{A nonequilibrium viewpoint}

The statistical behavior of the SSM is therefore neither purely equilibrium criticality nor purely stationary turbulence. It is better viewed as a nonequilibrium network of magnetic states whose coarse variables drift, synchronize, desynchronize, and occasionally reverse. In this picture, $1/f$-like PSDs and heavy-tailed amplitudes are distinct projections of a single evolving collective process.

\section{Unified Magnetic-Topological Model for Coronae, Flares, Winds, and Jets}

The same collective variables that control the timing statistics can also classify the outflow and dissipation channels.

\begin{figure}
    \centering
    \includegraphics[width=1\linewidth]{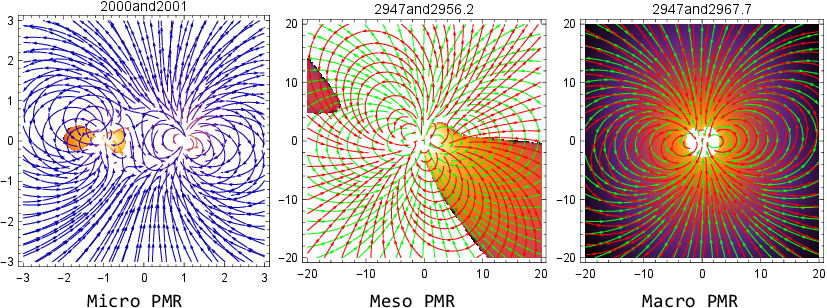}
    \caption{Magnetic-field comparison on the $y=0$ plane in the SSM simulation. The magnetic field at two nearby times is shown in different colors. Regions where the inner product of the two fields is negative are highlighted in orange or darker colors: the potential patch of magnetic reconnection and magnetic activity (PMR). 
    The magnetic field is calculated as follows. We set N-spins uniformly on the circle of radius 1 on the x-y plane centered at the origin. Assuming each spin is a magnetic moment, we superpose all the magnetic fields from the spins.  
    Left: \textbf{Micro PMR} The case of steady polarity. The reversed patch is small and local, but it persists at all times due to the persistent spin fluctuations.  
    Middle: \textbf{Meso PMR} The case of polarity excursion. The candidate zone of magnetic reconnection expands, yielding large active regions. 
    Right: \textbf{Macro PMR} The case of the full polarity flip. After a global polarity reversal, the field inner product is essentially negative everywhere, indicating a configuration in which magnetic energy can be efficiently converted into kinetic energy on large scales.}
    \label{fig:7}
\end{figure}


\subsection{Closed topology and coronae}

When the flux is predominantly closed, repeated local reconnection events heat the plasma without producing a strong organized outflow. Coronal luminosity will be expressed as 
\begin{equation}
L_{\rm cor}=A_c(1-f_{\rm open})\left[Q_0+Q_1\mathcal{G}\right],
\label{eq:Lcor}
\end{equation}
where $A_c$ is a geometric conversion factor, $Q_0$ is a background dissipation level, and $Q_1\mathcal{G}$ represents the reconnection-enhanced contribution. The factor $(1-f_{\rm open})$ suppresses the corona when a large fraction of the flux is open. Equation~(\ref{eq:Lcor}) is therefore the simplest closure that increases with closed flux and with magnetic inhomogeneity.

\begin{figure}
    \centering
    \includegraphics[width=0.75\linewidth]{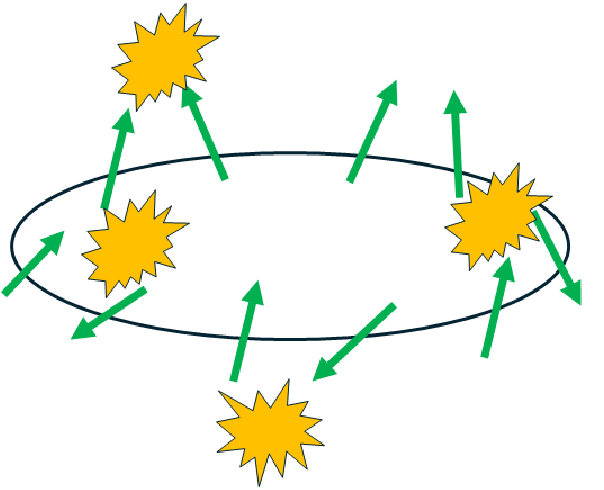}
    \caption{Schematic SSM picture of the corona. Coronal structures form above and below the region of the inner disk occupied by the coarse-grained spins. In the model, local and mesoscopic reconnection in this environment continuously heats the corona, whereas stronger large-scale ordering is required to form an axial jet channel.}
    \label{fig:8}
\end{figure}

Figure~\ref{fig:8} sketches the corresponding geometry: the corona forms above and below the region in the inner disk where the magnetic domains or spins are located.

\subsection{Partial coherent reversal and flares}

Flares are strongest when reconnection becomes coherent over a substantial fraction of the system without yet turning into a fully open axial structure. A simple phenomenological form of the flare power is
\begin{equation}
P_{\rm flare}=A_f\mathcal{G}\,\Theta(M-M_c),
\label{eq:Pflare}
\end{equation}
where $A_f$ is a normalization constant, $\Theta$ is the Heaviside step function, and $M_c$ is the minimum global order required for large coherent events. The logic is that strong local gradients alone are not enough: the largest flares are expected when there is both substantial reconnection activity and enough large-scale organization to couple many sites together.

\subsection{Partially open topology and winds}

Broad winds require open flux but not necessarily strong axial alignment. We therefore write the wind power as 
\begin{equation}
P_{\rm wind}=f_{\rm open}(1-C_{\rm ax}^2)\left[A_w\mathcal{S}_{\rm wind}(\dot m)+B_w\mathcal{G}\right],
\label{eq:Pwind}
\end{equation}
where $\dot m$ is the dimensionless accretion rate, $\mathcal{S}_{\rm wind}(\dot m)$ is a smooth source function for the quasi-steady wind background, and $B_w\mathcal{G}$ measures the additional clumpy or intermittent contribution from magnetic restructuring. The prefactor $f_{\rm open}(1-C_{\rm ax}^2)$ encodes the intended geometry: wind power grows with open flux but is suppressed when the open flux becomes too strongly axis-aligned, in which case the system prefers a jet-like channel instead.

\subsection{Open axial topology and jets}

Jets require both open flux and strong axial organization. A compact phenomenological closure for the jet power is
\begin{equation}
P_{\rm jet}=A_j f_{\rm open}C_{\rm ax}^2M^\alpha\mathcal{E}_{\rm rot}(\dot m)+B_j f_{\rm open}C_{\rm ax}^2\left|\frac{dM}{dt}\right|+C_j\Gamma_{\rm rev,glob},
\label{eq:Pjet}
\end{equation}
where $\mathcal{E}_{\rm rot}(\dot m)$ is an effective rotational energy supply function, $\alpha>0$ controls how steeply steady jet power grows with large-scale order, and $\Gamma_{\rm rev,glob}$ is a phenomenological source term associated with global reversal or excursion. The first term represents the steady jet channel sustained by ordered open flux. The second term captures transient enhancement during rapid magnetic restructuring. The last term allows an additional burst-like contribution when global topology changes coherently. Equation~(\ref{eq:Pjet}) is therefore the simplest additive split between steady and impulsive jet channels.

\section{Observational Connection I: statistical properties}
\label{sec:connectionI}
We now compare the statistical diagnostics of the SSM with long-term X-ray variability data.

\subsection{Main agreement: the demodulated envelope is the relevant variable}

The clearest agreement concerns the \emph{demodulated} signal rather than the raw one.
In the SSM example of Fig.~\ref{fig:5}, the detrended raw series is not pink, while the absolute value
of the detrended series becomes pink and follows Taylor's law. This is exactly the behavior
expected if synchronization first generates an amplitude envelope and the low-frequency
statistics are revealed only after a demodulation-like operation.

\begin{figure}
    \centering
    \includegraphics[width=1\linewidth]{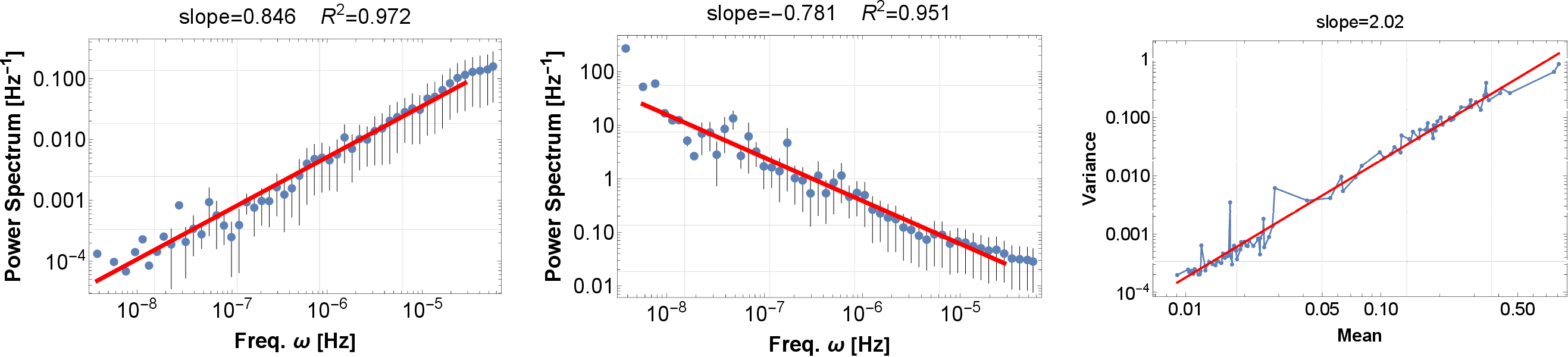}
    \caption{MAXI 16 yr diagnostics for GRS 1915+105, Galactic-Compact-Binary-BH, parallel to Fig.~\ref{fig:5}. 
    Left: PSD of the detrended light curve. 
    Middle: PSD of the absolute value of the detrended light curve, which becomes pink. 
    Right: Taylor’s law test. As in the SSM calculation, the demodulated envelope is the quantity that most clearly displays both pink noise and Taylor’s law behavior.}
    \label{fig:9}
\end{figure}

A similar pattern is seen observationally in at least part of the MAXI sample. Figure~\ref{fig:9} shows
that for GRS~1915+105 the detrended raw light curve is not pink, whereas the absolute value
of the detrended series becomes pink and satisfies Taylor's law. In this sense, the SSM captures
a nontrivial qualitative feature of the data: the relevant long-memory structure may be hidden
in the envelope and need not appear directly in the signed fluctuation itself.

\subsection{Main mismatch: SSM appears to overproduce this behavior}

The problem is that the present SSM tends to produce this transformation too broadly.
Even for parameter sets where the raw detrended SSM series is not pink, taking the absolute
value often makes it pink and Taylor-like. In the model, hidden-envelope statistics are therefore
rather generic.

The MAXI data are more selective. Empirically, sources that are already close to pink in the
original light curve often remain pink and Taylor-like after the absolute-value operation.
However, for many sources that are not pink to begin with, taking the absolute value does
\emph{not} produce pink noise or Taylor's law. Thus the observational class
``non-pink raw $\to$ pink absolute value'' certainly exists, but it is not nearly as universal
as in the present SSM.

\subsection{Interpretation}

This mismatch suggests that synchronization and nonlinear demodulation are important but not
sufficient. The current SSM captures the mechanism by which envelope extraction can reveal
hidden long-timescale organization, but it does not yet include enough of the additional physics
that determines whether such organization survives in the observed X-ray flux.

The most likely missing ingredients are straightforward. The observed flux is not the order
variable itself but a radiatively filtered observable. Real sources also contain additive
contamination from other variability channels, state mixing, and finite observational windows.
Any of these can suppress the pink/Taylor signature even if the underlying magnetic dynamics
contains an envelope hierarchy. In this sense, the present SSM is best viewed as describing one
important variability component rather than the full observational transfer problem.

These issues lead us to go beyond static diagnostics and investigate the detailed synchronization history across the entire observation time series. 

\begin{figure}
    \centering
    \includegraphics[width=0.8\linewidth]{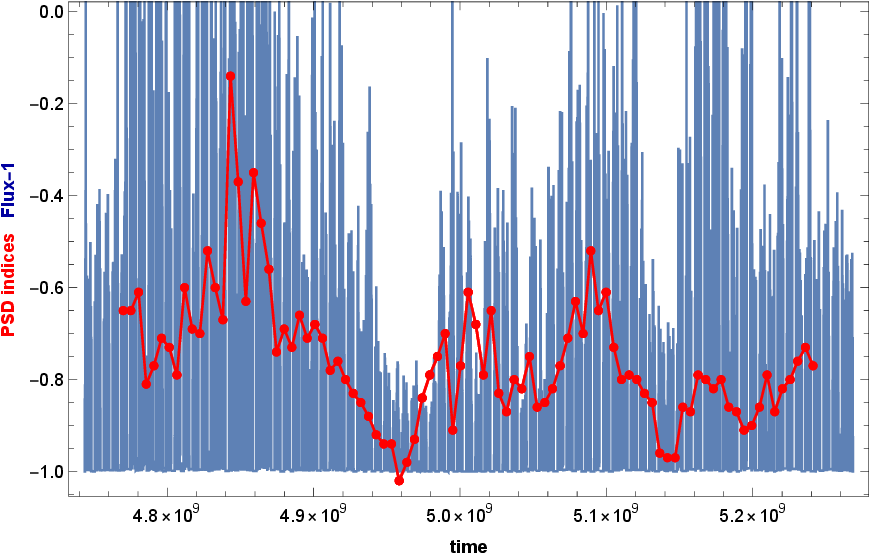}
    \caption{MAXI 16 yr running-index analysis for Cyg X-1, parallel to Fig.~\ref{fig:6}. 
    Shown are the normalized X-ray flux shifted downward by 1 (blue) and the running PSD indices (red) obtained using the same method as in Fig.~\ref{fig:6}.     
    The PSD slope tends to be steeper at lower X-ray intensity and flatter, closer to white noise, at higher intensity. This suggests that the degree of long-timescale synchronization changes across the observed variability cycle.}
    \label{fig:10}
\end{figure}

\subsection{Time-dependent variation of the PSD slope: comparison of Fig.~\ref{fig:6} and Fig.~\ref{fig:10}}

A further important point is the \emph{time dependence} of the PSD slope within a single data set.
The comparison between the SSM result in Fig.~\ref{fig:6} and the MAXI result in Fig.~\ref{fig:10} suggests that
the low-frequency PSD index is not a fixed property of a source, but varies in time together
with the degree of large-scale ordering.

Figure~\ref{fig:6} suggests that, within the SSM, the running PSD index changes systematically along the same time series as the
collective spin variable. In particular, during polarity reversals or large excursion-like episodes,
the PSD slope tends to become white (the slope shallower than pink), indicating that strong global rearrangement weakens
the frequency crowding and partial synchronization that otherwise support the low-frequency
continuum. Conversely, during more ordered epochs, the PSD is more likely to approach a pink
spectrum. Thus, the PSD index and the degree of collective spin ordering are dynamically linked
within one realization.

Figure~\ref{fig:10} suggests an observational counterpart of the same idea. In the MAXI light curve, the
running PSD index also changes within a single source over time, rather than remaining constant.
This indicates that the observed spectrum is evolving together with the source state. Although the observational quantity is not the spin order parameter itself, the qualitative similarity to
Fig.~\ref{fig:6} is suggestive: in both the model and the data, the spectral index moves as the internal
state of the system changes.


This dynamical viewpoint motivates the next section. If the PSD slope varies together with the
degree of ordering in both the SSM and the observed light curve, then the statistical properties
and the large-scale state evolution should be understood within a common framework rather than
as separate phenomena. In the following section, we therefore examine this evolving state
structure more explicitly.

\section{Observational Connection II: cycles}
\label{sec:cycles}

\subsection{Minimal three-level phenomenological variables for the q-diagram}
The micro/meso/macro PMR labels in Fig. \ref{fig:7} should be understood as a geometric illustration of the same multilevel viewpoint later encoded phenomenologically by the reduced variables \(m(t),s(t),M(t)\), rather than as a one-to-one definition of those variables.

\begin{figure}
    \centering
    \includegraphics[width=0.7\linewidth]{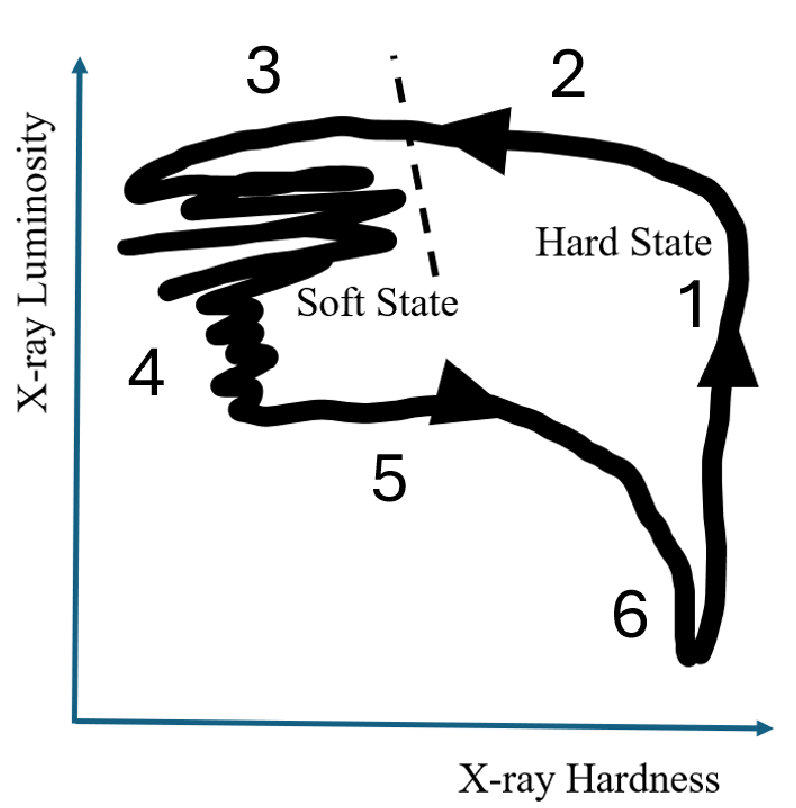}
    \caption{Canonical hardness--intensity diagram (q-diagram) of black-hole X-ray binaries. The SSM interpretation assigns the hard branch (1) to a noisy but globally organized magnetic state, through the intermediate state (2), the jet-line region (3) to rapid large-scale magnetic restructuring with strong transient ejecta, and the soft branch (4) to a thermally stable but less dipolar magnetic configuration with suppressed steady jet power. Through quiet transient states (5-6), the system returns to the hard state (1). }
    \label{fig:11}
\end{figure}

Figure~\ref{fig:11} summarizes the phenomenological state cycle, or q-diagram, to be interpreted. The SSM reading is that hard states retain comparatively strong macro-order and open axial flux, the jet-line region corresponds to rapid restructuring with strong transient ejecta, and soft states suppress the steady jet because the global dipole-like order is weaker even though local magnetic activity remains present.

To formalize this picture, we introduce four reduced variables,
\begin{align}
m(t) &\ge 0 && \text{: micro-scale fluctuation intensity: level-1=micro}, \nonumber\\
s(t) &\in [0,1] && \text{: meso-scale synchronization or clustering: level-2=meso}, \nonumber\\
M(t) &\in [-1,1] && \text{: signed macro-scale dipolar order parameter: level-3=macro}, \nonumber\\
H(t) &\ge 0 && \text{: history variable storing helicity, free magnetic energy, or topology memory}.\nonumber
\end{align}
The purpose of this set is not uniqueness, but economy: $m$ captures fast variability, $s$ captures intermediate coherence, $M$ captures large-scale polarity and order, and $H$ introduces hysteresis.

\subsection{Minimal dynamical system}

The following equations are not derived from first-principles GRMHD. They are a reduced phenomenological model designed to capture the observed state sequence, hysteresis, and the distinction between steady and transient jet channels.

\paragraph{Micro-scale activity}
\begin{equation}
\frac{dm}{dt}
=
\epsilon_0
+\epsilon_1 \left(\frac{dM}{dt}\right)^2
+\epsilon_2 s
-\gamma_m m.
\label{eq:micro_eq}
\end{equation}
Here $\epsilon_0$ is a baseline fluctuation source, $\epsilon_1(dM/dt)^2$ expresses additional agitation during rapid macro-scale restructuring, $\epsilon_2 s$ allows meso-scale synchronization to feed smaller-scale variability, and $\gamma_m$ is a decay rate.

\paragraph{Meso-scale synchronization}
\begin{equation}
\frac{ds}{dt}
=
\alpha_s\, m(1-s)
+\beta_s |M|(1-s)
-\chi_s s
-\eta_s \left(\frac{dM}{dt}\right)^2 s.
\label{eq:meso_eq}
\end{equation}
The two growth terms encode the idea that both micro-scale stirring and existing macro-order can seed meso-scale clustering, while the factors $(1-s)$ saturate the growth as $s\to1$. The term $\chi_s s$ is a linear loss term, and $\eta_s(dM/dt)^2 s$ suppresses meso-scale coherence during violent global rearrangement.

\paragraph{Macro-scale dipolar order}
\begin{equation}
\frac{dM}{dt}
=
a(H,m)\, M
-b M^3
+c\, s
-d\, \Xi \, M
+\xi_M(t),
\label{eq:macro_eq}
\end{equation}
with
\begin{equation}
a(H,m)=a_0+a_H H-a_m m.
\label{eq:a_coeff}
\end{equation}
Here, $a_0$ is the baseline linear growth coefficient, $a_HH$ lets stored history stabilize large-scale order, and $a_mm$ lets strong micro-turbulence weaken it. The cubic term $-bM^3$ is the standard Landau saturation that prevents unbounded growth. The term $cs$ says that meso-scale synchronization can seed macro-order. The factor $d\Xi M$ allows accretion geometry or thermodynamic state, summarized by $\Xi$, to suppress or reshape the large-scale field. Finally, $\xi_M(t)$ is a stochastic forcing term.

\paragraph{History variable and hysteresis}
\begin{equation}
\frac{dH}{dt}
=
\mu_H |M| s
-\nu_H \left(\frac{dM}{dt}\right)^2 H
-\gamma_H H.
\label{eq:history_eq}
\end{equation}
The source term $\mu_H|M|s$ states that helicity or free magnetic energy is stored when meso- and macro-order coexist. The term $\nu_H(dM/dt)^2H$ releases that stored energy during rapid reorganization, while $\gamma_HH$ describes slow relaxation. This is the minimal mechanism by which hysteresis enters the q-diagram interpretation.



\subsection{State-by-state interpretation of the q-diagram}

The resulting interpretation is summarized in Table~\ref{tab:qdiagram_ssm}. The hard state (1) is noisy but globally ordered; the jet-line region (3) is a restructuring zone with large $|dM/dt|$; the soft state (4) is thermally stable but magnetically less dipolar; and the decay branch (5) is not the mirror image of the rise because the history variable $H$ produces hysteresis.

\begin{sidewaystable}[p]
\centering
\caption{Three-level SSM interpretation of the black-hole X-ray binary q-diagram.
\cite{FenderBelloniGallo2004, Belloni2010, IngramMotta2019, Fender2009Testing}
}
\label{tab:qdiagram_ssm}
\renewcommand{\arraystretch}{1.08}
\setlength{\tabcolsep}{4pt}
\footnotesize
\begin{tabularx}{\textheight}{>{\raggedright\arraybackslash}p{0.04\textheight} >{\raggedright\arraybackslash}p{0.09\textheight} >{\raggedright\arraybackslash}p{0.17\textheight} >{\raggedright\arraybackslash}p{0.18\textheight} >{\raggedright\arraybackslash}p{0.11\textheight} >{\raggedright\arraybackslash}p{0.15\textheight} >{\raggedright\arraybackslash}p{0.17\textheight}}
\toprule
\textbf{No.} & \textbf{State} & \textbf{Observed phenomenology} & \textbf{Three-level SSM picture} & \textbf{Typical ordering} & \textbf{Consistency / limits} & \textbf{SSM predictions} \\
\midrule
1 & Hard state & Strong broad-band noise; high RMS; compact flat-spectrum jet; radio/IR/X-ray correlation. & Noisy but globally ordered. Net poloidal flux sustains a steady jet while micro-fluctuations remain strong. & $m$ high; $s$ med--high; $|M|$ high; $H$ building. & Jet-producing state with persistent variability; not full synchrony. & Mean polarization axis is stable with rapid swings. 
\\
\addlinespace
2 & HIMS (rise) & Disk moves inward; 
IR drops; radio becomes variable and optically thin. & Global order weakens; meso-clusters become fragile; system approaches large-scale reorganization. & $m$ high; $s$ high but unstable; $|M|$ falling; $H$ large. & 
Jet instability starts. & Non-stationarity increases in lags and polarization.
\\
\addlinespace
3 & Jet line / SIMS & Fast ejecta; transient radio flare; RMS drop.
 & Rapid magnetic restructuring: reversal or strong excursion releases stored energy and drives discrete ejecta. & $m$ high; $s$ rapidly varying; $M$ rapidly varying; $|dM/dt|$ maximal; $H$ released. & Best SSM match; full reversal is possible but not required. & Rapid polarization-angle swings; timing anomalies; knotty/two-sided ejecta; flare strength tracks $|dM/dt|$ more than $|M|$. \\
\addlinespace
4 & Soft state & Thermal disk to ISCO; steady jet quenched; 
weak radio may persist early; non-thermal tail may remain. & Thermally stable but weakly dipolar. Local reconnection continues, yet no strong jet-supporting macro field survives. & $m$ high; $s$ low--med; $M$ low; $H$ relaxing. & Explains jet quenching without removing all magnetic activity. & Local flares and coronal tails remain possible, but a sustained compact jet is hard to maintain. \\
\addlinespace
5 & Return HIMS (decay) & Similar to upper HIMS, but without strong optically thin flare; radio recovers at lower luminosity. & Large-scale order is rebuilt gradually rather than explosively. & $m$ high; $s$ med; $M$ rising; $H$ still important. & Hysteresis follows naturally if $H$ stores topology/free energy. & Rise and decay should differ in polarization, 
and radio/X-ray correlation at the same hardness. \\
\addlinespace
6 & Quiescence & Low luminosity; recessed disk; weak timing activity; UV/optical thermal component. & Weakly excited, weakly coupled state; only low-level micro-activity remains. & $m$ low--med; $s$ low; $M$ low; $H$ fading. & Qualitatively consistent; detailed quiescent flow is better modeled in RIAF/evaporation pictures. & 
\\
\bottomrule
\end{tabularx}

\vspace{0.4em}
\begin{minipage}{0.95\textheight}
\footnotesize
\textit{Abbreviations:} ISCO, innermost stable circular orbit; IR, infrared.
\end{minipage}
\end{sidewaystable}

\section{Quantitative tests and observational falsifiability of the SSM scenario}
\label{sec:ssm_tests}

In this section, we present the main order-of-magnitude checks and falsifiable predictions of the synchronized spin model (SSM) for cosmic magnetic activity. The basic viewpoint is that the dominant energy reservoir is the gravitational potential energy released by accretion, while magnetic fields store, redistribute, and intermittently release a fraction of this energy through reconnection events on various scales. 
Thus, the SSM yields quantitative expectations for the energetics of discrete jet knots, the hierarchical relation among jet--flare--corona phenomena, alternating knot amplitudes, magnetic polarity relics, and the response of AGN luminosity to rapid changes in the accretion state \cite{Yamada2010,BlandfordZnajek1977,Morokuma2025}.

\subsection{Energetic consistency from discrete jets to the corona}
\label{subsec:ssm_energetics_final}

\subsection*{(i) knot energy budget}

As a fiducial upper-envelope estimate, let us denote by $E_{B,\max}$ the magnetic energy
that can participate in one major knot-launching episode associated with a large-scale reversal
or excursion. The important point is that this should not be interpreted as the instantaneous
magnetic energy of a single compact blob of size $\sim 100\,r_g$. Observational and theoretical
studies of AGN jets increasingly indicate that acceleration and collimation proceed over an
extended acceleration--collimation zone (ACZ), often reaching $\lesssim 10^5$--$10^6\,r_g$
from the black hole \cite{Yi2024ACZ,Hada2024M87Review}. We therefore adopt a more realistic
effective scale,
\begin{equation}
B \sim 300~{\rm G}, \qquad R_{\rm eff}\sim 2000\,r_g,
\end{equation}
for the inner magnetically active part of the ACZ that may participate coherently in a single
major release episode. This choice is still conservative in the sense that $R_{\rm eff}$ is far
smaller than the full observed ACZ extent.

With this interpretation, the available magnetic energy scales as
\begin{equation}
E_{B,\max}\sim \frac{B^2}{8\pi}\,\frac{4\pi}{3}R_{\rm eff}^3
\propto B^2 R_{\rm eff}^3 ,
\end{equation}
and the above setting 
yields the order estimate, 
\begin{equation}
E_{B,\max}\sim {\rm few}\times 10^{53}\text{--}10^{54}~{\rm erg},
\end{equation}
which is the scale relevant for the largest AGN knot-launching episodes discussed below.
In this reading, $E_{B,\max}$ represents the effective magnetic reservoir of the inner ACZ
participating in one major event, rather than the magnetic content of a single compact plasmon.

Considering the efficiency rate $\eta_k$ for the conversion of magnetic energy into the bulk
energy of a discrete knot, the ejecta energy is
\begin{equation}
E_{\rm knot}\simeq \eta_{\rm rec}\eta_{\rm bulk}E_B \equiv \eta_k E_B,
\qquad 0<\eta_k<1.
\end{equation}
The remainder is naturally expected to be partitioned into thermal energy, non-thermal
particles, radiation, and residual Poynting flux \cite{Yamada2010,Kino2015}.

If the knot is ejected with Lorentz factor $\Gamma$, its bulk kinetic energy is
\begin{equation}
E_{\rm knot}=(\Gamma-1)M_{\rm knot}c^2.
\end{equation}
Using the fiducial upper-envelope value above, one obtains
\begin{equation}
M_{\rm knot}\simeq 1.0\,\eta_k
\left(\frac{E_B}{8\times10^{53}~{\rm erg}}\right)
\left(\frac{\Gamma-1}{0.45}\right)^{-1} M_\odot,
\end{equation}
where $\Gamma\simeq 1.45$ corresponds to $\beta\simeq 0.7$. Thus, for a mildly relativistic
plasmon, a knot mass of order $M_{\rm knot}\sim M_\odot$ is not inconsistent with the
fiducial upper-envelope estimate. 
For faster knots, the implied mass decreases rapidly,
\(M_{\rm knot}\sim 0.1\,\eta_k\,M_\odot \quad (\Gamma\simeq 5.5,\ \beta\simeq 0.98),\) and 
\(M_{\rm knot}\sim 0.05\,\eta_k\,M_\odot \quad (\Gamma\simeq 10,\ \beta\simeq 0.995).\)
Therefore, the SSM does not require a unique knot mass. 

\subsection*{(ii) hierarchical energy-per-decade}

A useful extension of this argument is obtained if a suitably defined catalog of reconnection-driven events obeys an approximate occurrence distribution
\begin{equation}
\frac{dN}{dE}=AE^{-\alpha},\qquad \alpha\simeq2,
\end{equation}
as often discussed for flare-like avalanches and SOC-like systems \cite{Aschwanden2011SOC,Aschwanden2025}.
Here, the key point is interpretive: this form is not assumed to describe the full continuous flux
distribution of AGN/XRB light curves, which is often closer to log-normal in the multiplicative-
variability picture \cite{Uttley2005}. Instead, it is used as an effective event-catalog model
for the subset of reconnection-driven bursts that can be identified as discrete flares, shots, or
ejecta. In solar and stellar flare systems, this approximation is often directly useful, while in
black-hole systems, it should be regarded as an illustrative hypothesis whose validity depends on
event definition and source state \cite{Maehara2012,Maehara2015}.

The energy released \(W\) per logarithmic interval is then
\begin{equation}
\frac{dW}{d\ln E}
=E\frac{dN}{d\ln E}
=E^2\frac{dN}{dE}
=AE^{2-\alpha}.
\end{equation}
For $\alpha=2$, this becomes
\begin{equation}
\frac{dW}{d\ln E}=A={\rm const.},
\end{equation}
so that each decade or octave in event energy carries approximately the same total released
energy. In the SSM context, this provides a convenient energetic bridge between the largest
identified events (discrete jet ejections), intermediate events (large flares or clumpy wind
ejecta), and the smallest identified events (unresolved reconnection bursts contributing to
coronal heating). 

Here $E_{\max}\sim10^{54}\,{\rm erg}$ should be understood as the upper end of the effective
event hierarchy associated with major ACZ-scale magnetic episodes, not as the energy of a
single elementary reconnection site. If $E_{\max}\sim10^{54}$ erg and $E_{\min}\sim10^{42}$ erg,
the hierarchy spans
\(\log_{10}\!\left({E_{\max}}/{E_{\min}}\right)\sim 12\)
decades, and therefore the average power deposited per decade is roughly
\begin{equation}
L_{\rm per~decade}\sim
\frac{L_{\rm mag}}{\log_{10}(E_{\max}/E_{\min})}
\sim
\frac{L_{\rm mag}}{12}, 
\end{equation}
where $L_{\rm mag}$ is the total luminosity associated with magnetic activity.
This means that, once the magnetic budget is sufficient to power rare giant knot-launching
events, it is energetically plausible that lower-energy bands of the same hierarchy can sustain
persistent flaring and coronal heating.

A convenient way to interpret the power scale used below is to convert the largest knot-
launching event into a time-averaged magnetic power per logarithmic energy interval. If the
upper-end episode has characteristic energy $E_{\max}\sim10^{54}\ {\rm erg}$ spanning about one decade, and occurs once
per $\Delta t_{\max}\sim300\ {\rm yr}\sim10^{10}\ {\rm s}$, then
\begin{equation}
L_{\rm per~decade}\sim \frac{E_{\max}}{\Delta t_{\max}}
\sim10^{44}\ {\rm erg\ s^{-1}},
\end{equation}
suggesting comparable power per logarithmic interval across the jet, wind, and coronal bands, if such a hierarchy is realized. 

\subsection*{(iii) corona/wind consequences}

A simple consistency check of coronal-temperature \(T_{\rm cor}\), coronal-Luminosity \(L_{\rm cor}\), and coronal-energy \(U_{\rm cor}\) follows from
\begin{equation}
U_{\rm cor}\sim 3n_{\rm cor}V_{\rm cor}k_{\rm B}T_{\rm cor},
\end{equation}
together with
\begin{equation}
U_{\rm cor}\sim L_{\rm cor}t_{\rm res},
\end{equation}
where $n_{\rm cor}$, $V_{\rm cor}$, and $t_{\rm res}$ are the coronal density, volume, and effective
energy residence time. One then finds
\begin{equation}
T_{\rm cor}\sim
\frac{L_{\rm cor}t_{\rm res}}
{3n_{\rm cor}V_{\rm cor}k_{\rm B}}.
\end{equation}
If, owing to the approximately equal-energy-per-decade property, the coronal heating band
receives a fraction $\epsilon_{\rm cor}$ of the total magnetic power, $L_{\rm cor}\sim\epsilon_{\rm cor}L_{\rm mag}$,
then
\begin{equation}
T_{\rm cor}\sim 7.2\times10^8~{\rm K}
\left(\frac{\epsilon_{\rm cor}L_{\rm mag}}{10^{44}~{\rm erg\ s^{-1}}}\right)
\left(\frac{t_{\rm res}}{10^{4}~{\rm s}}\right)
\left(\frac{n_{\rm cor}}{10^{9}~{\rm cm^{-3}}}\right)^{-1}
\left(\frac{V_{\rm cor}}{10^{46}~{\rm cm^{3}}}\right)^{-1}.
\end{equation}
This is deliberately a one-zone upper-envelope estimate, but it shows that X-ray-coronal
temperatures of order $10^{8}$--$10^{9}$ K are not energetically problematic if only a modest
fraction of the SSM magnetic budget is thermalized in a compact corona. Here
$t_{\rm res}\sim10^{4}$ s is adopted as a fiducial residence time appropriate for a compact AGN
corona, corresponding to an order-of-magnitude light-crossing / escape / dissipation timescale
of a region of several to several tens of $r_g$ around a $10^8\,M_\odot$ black hole.

A similar order-of-magnitude estimate may be made for the wind velocity if a fraction
$\epsilon_{\rm w}$ of the magnetic power is deposited into clumpy wind ejecta rather than into
the corona or the jet. Writing the wind kinetic power as
\begin{equation}
L_{\rm w}\sim \frac{1}{2}\dot{M}_{\rm w}v_{\rm w}^{2},
\end{equation}
one obtains
\begin{equation}
v_{\rm w}\sim
\left(\frac{2L_{\rm w}}{\dot{M}_{\rm w}}\right)^{1/2}
\simeq 4.0\times10^{8}
\left(\frac{L_{\rm w}}{10^{43}~{\rm erg\ s^{-1}}}\right)^{1/2}
\left(\frac{\dot{M}_{\rm w}}{1~M_\odot~{\rm yr^{-1}}}\right)^{-1/2}
{\rm cm\ s^{-1}}.
\end{equation}
Thus, even for a moderate mass-loading rate, reconnection-powered wind ejecta can naturally
reach velocities of order $10^{3}$--$10^{4}\ {\rm km\ s^{-1}}$. For smaller $\dot{M}_{\rm w}$ or more
concentrated energy deposition, the same estimate extends into the ultra-fast outflow regime,
$v_{\rm w}\sim0.1c$, consistent with observed AGN ultra-fast outflows and with the recent
XRISM evidence that at least some quasar winds are highly structured rather than smooth and
continuous \cite{Tombesi2010,XRISM2025NatureStructuredWinds}. This suggests that
at least part of the observed AGN wind phenomenology may be better viewed not as a perfectly
continuous flow, but as a hierarchy of discrete, bullet-like ejecta driven by multiscale magnetic
reconnection.

In the SSM picture, such wind bullets occupy an intermediate part of the same reconnection
hierarchy that connects the largest jet-launching events to the smallest coronal-heating events.
By contrast, if the relevant event-energy distribution is closer to log-normal,
\begin{equation}
p(E)=\frac{1}{E\sigma\sqrt{2\pi}}
\exp\!\left[-\frac{(\ln E-\mu)^2}{2\sigma^2}\right],
\end{equation}
then the energy released per logarithmic interval becomes
\begin{equation}
\frac{dW}{d\ln E}\propto E^2p(E),
\end{equation}
which peaks around a characteristic scale $E_*=\exp(\mu+\sigma^2)$ rather than remaining flat
across decades. Hence, the equal-energy-per-decade argument is specific to the effective power-law
event-catalog approximation and should not be confused with the log-normal description of the
continuous multiplicative variability.

\subsection{Alternating knot amplitudes and the doubled cycle}
\label{subsec:ssm_alternation}

A distinctive SSM prediction is that if a slowly varying large-scale background field $B_g$
coexists with a polarity-reversing dynamo component $B_d$, then successive knot-launching
events need not be energetically identical. In the simplest picture, the reconnecting field
alternates between constructive and destructive combinations,
\begin{equation}
B_{{\rm rec},\pm} \sim B_d \pm B_g ,
\end{equation}
so that the large-to-small knot energy ratio is estimated as
\begin{equation}
\frac{E_{\rm large}}{E_{\rm small}}
=
\left(
\frac{B_d + B_g}{B_d - B_g}
\right)^2 .
\end{equation}
For example,
\begin{align}
\frac{B_{\rm g}}{B_{\rm d}}=0.1
\quad &\Rightarrow \quad
\frac{E_{\rm large}}{E_{\rm small}}
=
\left(\frac{1.1}{0.9}\right)^2
\simeq 1.49,\\
\frac{B_{\rm g}}{B_{\rm d}}=0.2
\quad &\Rightarrow \quad
\frac{E_{\rm large}}{E_{\rm small}}
=
\left(\frac{1.2}{0.8}\right)^2
=
2.25.
\end{align}
Thus, even a background field at the 10--20\% level of the reversing field can generate a robust factor-of-$1.5$--$2$ alternating modulation in successive knot amplitudes. Observationally, one may search for this by fitting the knot flux sequence $F_n$ with
\begin{equation}
F_n = \bar F \left[1+\epsilon\cos(\pi n+\phi)\right],
\label{eq:Fn_fit}
\end{equation}
or equivalently by testing whether neighboring ratios $R_n\equiv F_{n+1}/F_n$ exhibit a statistically significant even--odd asymmetry. In this picture, the knot spacing traces the fundamental reversal cadence, while the brightness alternation traces the doubled cycle.

A complementary timing estimate comes from the deprojected knot spacing $\Delta r$ and the intrinsic knot speed $\beta c$:
\begin{equation}
T_{\rm ej}\sim \frac{\Delta r}{\beta c}.
\label{eq:Tej}
\end{equation}
For the quasi-periodic knot train in PKS~0637$-$752, the observed projected separation is of order $\Delta r_{\rm proj}\sim 7.6~{\rm kpc}$ \cite{Godfrey2012}. After deprojection, the inferred modulation timescale is of order $10^{3}$--$10^{5}$~yr, depending on the jet speed and geometry \cite{Godfrey2012}. Within the SSM, this is interpreted as an estimate of the large-scale magnetic reversal timescale. The same logic can be applied to systems of very different masses by using the dimensionless ratio
\begin{equation}
\frac{T_{\rm ej}}{t_g},
\qquad
t_g\equiv \frac{GM_\bullet}{c^3},
\label{eq:Tgtg}
\end{equation}
which provides a direct route to testing whether protostellar jets, microquasars, and AGN jets lie on a common hierarchy when measured in gravitational time units.

\subsection{Magnetic polarity relics in the circumnuclear medium}
\label{subsec:ssm_polarity_relics}

If a sequence of SSM reversals repeatedly injects magnetic flux of alternating polarity into the surrounding medium, the resulting field geometry may leave a spatially stratified relic pattern, provided that magnetic advection dominates over turbulent diffusion on the corresponding scale. The characteristic width of one polarity domain is
\begin{equation}
\lambda_{\rm rev}\sim v_{\rm adv}T_{\rm rev},
\label{eq:lambdarev}
\end{equation}
where $v_{\rm adv}$ is an effective transport speed and $T_{\rm rev}$ is the reversal period. Adopting representative nuclear values motivated by \cite{LamastraEtAl2016NGC1068},
\begin{align}
v_{\rm adv} &\sim 100~{\rm km\,s^{-1}},\\
T_{\rm rev} &\sim 10^{4}~{\rm yr},
\end{align}
one finds
\begin{equation}
\lambda_{\rm rev}
\sim
(10^{7}\ {\rm cm\,s^{-1}})
(3.15\times10^{11}\ {\rm s})
\sim 3\times10^{18}\ {\rm cm}
\sim 1~{\rm pc}.
\label{eq:lambdarev_num1}
\end{equation}
For $T_{\rm rev}\sim 10^{5}$~yr, this becomes
\begin{equation}
\lambda_{\rm rev}\sim 10~{\rm pc},
\label{eq:lambdarev_num2}
\end{equation}
and somewhat larger values follow if $v_{\rm adv}\sim 300~{\rm km\,s^{-1}}$. Therefore, the most promising search region is not the entire galactic disk, but the central parsec-to-tens-of-parsecs environment, where alternating sign changes in polarization or Faraday rotation may survive as a fossil record of nuclear magnetic reversals. Mapping of magnetic-field structure in the Galactic environment already shows that coherent field reversals can be observationally reconstructed \cite{Doi2024}.

The competing process is turbulent diffusion, with a characteristic timescale
\begin{equation}
t_{\rm diff}\sim \frac{L^2}{\eta_{\rm turb}} .
\label{eq:tdiff}
\end{equation}
Using a fiducial turbulent diffusivity consistent with \cite{BrandenburgNtormousi2023GalacticDynamos},
\begin{equation}
L\sim 100~{\rm pc},
\qquad
\eta_{\rm turb}\sim 10^{26}\ {\rm cm^2\,s^{-1}},
\end{equation}
one obtains
\begin{equation}
t_{\rm diff}
\sim
\frac{(3\times10^{20}\ {\rm cm})^2}{10^{26}\ {\rm cm^2\,s^{-1}}}
\sim
10^{15}\ {\rm s}
\sim
3\times10^{7}\ {\rm yr}.
\label{eq:tdiff_num}
\end{equation}
Hence, parsec-scale to sub-kiloparsec-scale polarity relics are, at least in principle, long-lived enough to be observable, whereas clean signatures over full spiral-arm scales are expected to be more easily erased by galactic shear and turbulence.

The tomography of the Sagittarius arm \cite{Doi2024} shows that pc-scale coherent magnetic patches are observationally plausible, with smooth field structure on scales below \(\sim 10\,pc\). However, those clouds lie at Galactocentric radii of \(\sim 6\text{--}7\,kpc\), so they should be regarded as an existence proof of such coherence rather than as a direct test of the nuclear fossil-polarity scenario proposed here.

\subsection{Rapid AGN fading and the immediate energy source}
\label{subsec:ssm_fading}

An interesting observational test is provided by AGN that fade dramatically on
decadal timescales. The quasar SDSS J021801.90$-$003657.7, for example, showed a decline
from SDSS to HSC photometry by factors of order $10$--$20$ in the optical over roughly two
decades \cite{Morokuma2025,KomossaEtAl2025ExtremesAGN}. If the instantaneous radiative output $L_{\rm rad}$ is
controlled primarily by accretion power, one expects
\begin{equation}
L_{\rm rad}\simeq \eta \dot{M}c^2,
\end{equation}
with radiative efficiency $\eta\sim0.1$ in a standard disk. A luminosity decline by a factor
$f$ then corresponds, at the crudest level, to
\begin{equation}
\dot{M}_f \simeq \frac{\dot{M}_i}{f}.
\end{equation}

This argument is not unique to SSM: even in more standard accretion-disk pictures, a reduced
mass supply to the inner disk naturally leads to a long-lived decline of activity. The point of
the present example is therefore narrower. Such fading events support the general premise that
the immediate power source of observed AGN activity is the gravitational energy released by
accretion, not an order-unity change in the black-hole spin reservoir on decadal timescales.

What is more specific to the SSM picture is how the activity channels are reorganized during
that decline. Since SSM links coronae, flares, winds, and jets to a common evolving magnetic
state, the more discriminating test is whether these channels vary independently or in a
correlated manner during a fading event. In this sense, the observational target is not the
luminosity drop alone, but the pattern of channel rearrangement: whether coronal activity,
flare occurrence, wind signatures, and jet/non-jet behavior are reorganized coherently as the
system moves through magnetic state space.

\subsection{Additional falsifiable diagnostics}
\label{subsec:ssm_more_tests}

Beyond the tests discussed above, the SSM yields at least three further semi-quantitative diagnostics.

First, SSM predicts a symmetric bipolar ejection in both directions simultaneously associated with a polarity flip event. Thus, it predicts that the initial energies of the two opposite knots should satisfy
\begin{equation}
E_{+}\simeq E_{-},
\label{eq:bilateral}
\end{equation}
up to projection and Doppler effects. Hence, after correcting for beaming as far as possible, the intrinsic energetics of the two sides should be more symmetric than in models where the observed knot structure is generated mainly by environmental shocks. Twin-jet systems such as NGC~1052 provide useful laboratories for this check \cite{Vermeulen2003}.

Second, if large flares and knot ejections are different manifestations of the same reversal-driven reconnection event, the appearance of a new knot should follow a major flare after a propagation delay of order
\begin{equation}
\Delta t_{\rm flare\rightarrow knot}
\sim
\frac{r_{\tau\sim 1}}{\beta c},
\label{eq:delay}
\end{equation}
where $r_{\tau\sim 1}$ is the photospheric or radio-core distance at which the new knot becomes visible. Simultaneous multi-wavelength monitoring and VLBI imaging can test this directly.

Third, the jet width and collimation profile carry an imprint of how magnetically organized the outflow remains after launch. The observed collimation properties of systems such as 3C~273 \cite{Okino2022} provide a geometric constraint complementary to energetics: if the knot train is truly a sequence of compact magnetized plasmons rather than a quasi-continuous fluid, then long-distance collimation becomes less surprising, because each knot transports its own internal magnetic and kinetic structure.

Taken together, these tests show that the SSM is falsifiable in several independent ways. The model is supported if one finds: (i) knot-launching energetics compatible with a reconnection-powered magnetic reservoir; (ii) a roughly equal-energy hierarchy across jet, flare, and coronal bands when analyzed per logarithmic interval; (iii) an even--odd modulation of successive knot amplitudes consistent with a doubled cycle; (iv) parsec-scale relic polarity domains around nuclei; and (v) rapid AGN fading events whose energetics track accretion-state changes rather than changes in black-hole spin energy. Conversely, the model would be disfavored if high-quality data were to show that knot sequences are purely stochastic with no doubled-cycle component, that no bipolar energetic symmetry exists even after deprojection, or that the full AGN power output remains unchanged despite order-of-magnitude changes in the inferred inner accretion rate.


\section{Discussion}
We briefly outline SSM's niche within existing theories in this field. 

\begin{enumerate}
\item{What SSM adds}

SSM is not a replacement for GRMHD, radiative transfer, BZ/BP launching, or reconnection microphysics. Its role is narrower: it provides a mesoscopic state language that links variability, partial coherence, magnetic topology, and outflow morphology within one reduced variable set.

\item{Relation to existing frameworks}

Propagating-fluctuation models explain broadband timing; 
reconnection or plasmoid models explain impulsive dissipation; BZ/BP frameworks explain launching once an ordered geometry exists. The niche of SSM is to describe how the magnetic state moves between those geometries and why timing and morphology change together.

\item{Limitations}

The mapping from GRMHD fields to macro-spin domains remains phenomenological, the effective coefficients are not derived from first principles, and radiative and thermal physics are compressed into source functions. The present framework should therefore be regarded as a reduced state theory rather than as a source-by-source predictive simulation model.

\item{A possible validation route for the SSM}

The SSM is intentionally simple and phenomenological, and it is not intended as a direct derivation from GRMHD or as a source-by-source predictive model. Its value lies instead in providing a compact description of collective magnetic behavior in rotating conducting fluids. From this perspective, a possible route to validation is to ask whether similar multilevel magnetic activity can be identified across a wider range of cosmic systems, including laboratory dynamos \cite{Monchaux2007PRL,Berhanu2007EPL,Giesecke2010PRL}. 

The SSM suggests similarities between apparently different systems. For example, the q-diagram
in XRBs and the solar cycle may be viewed as different realizations of synchronize--desynchronize
cycles, and the shift of solar-flare PSD indices from pink toward whiter spectra near solar maximum
suggests the same link between spectral slope and global magnetic reorganization. At the laboratory
scale, the VKS (von K\'arm\'an sodium) experiment provides an instructive analog: a turbulent
rotating flow of liquid sodium exhibits self-excited dynamo action, polarity reversals, and excursions.
In the present language, these may be interpreted as emergent collective transitions of coarse-grained
local dynamo elements, although in VKS, the influence of boundary conditions, especially soft-iron
impellers, is known to be essential. Thus SSM is not an alternative to the low-dimensional mode
descriptions of VKS, but may provide a mesoscopic viewpoint underlying them. Table~2 sketches
a provisional extension of the same SSM language to other cosmic magnetic-activity systems and
to laboratory dynamos.

\begin{sidewaystable*}[p]
\centering
\caption{Provisional extension of the Synchronized Spin Model (SSM) to a broader range of cosmic magnetic-activity systems and to laboratory dynamos. This table is intended as a phenomenological unification map. Recurrent activity is common, but a strictly periodic global cycle analogous to the solar cycle is not established in all systems.}
\label{tab:ssm_unification_extended}
\setlength{\tabcolsep}{4.0pt}
\renewcommand{\arraystretch}{1.08}
\resizebox{\textheight}{!}{%
\begin{tabular}{p{2.7cm} p{2.5cm} p{2.4cm} p{2.8cm} p{4.0cm} p{2.7cm} p{5.0cm}}
\hline
System & Corona / magnetosphere & Flare / burst & Jet / CME / ejecta & Activity cycle or recurrent change & 1/f-like variability & Main driver in SSM language \\
\hline
XRB (BH) 
\cite{FenderBelloniGallo2004,Belloni2010,Uttley2005,
Gleissner2004CygX1}
& hot corona ($\sim 100$ keV)
& X-ray flare
& relativistic jet
& hard/soft cycle; hysteretic state transitions
& strong
& MRI-driven disk dynamo + multiscale reconnection \\
AGN \cite{LaMassa2015,MacLeod2016,KingPounds2015,Tombesi2010,Fukumura2017}
& X-ray corona
& X-ray flare
& relativistic jet / disk wind
& changing-look transition; long-term state change
& strong
& strong disk dynamo + flux accumulation / reorganization \\
NS (magnetar / high-B NS)
& magnetosphere
& X-ray / $\gamma$-ray flare, burst storm
& relativistic outflow / plasmoid-like ejecta
& recurrent outbursts; glitches may accompany activity, but no universal cycle
& present in some cases
& crust--core stress + magnetic twist / reconnection \\
Protostar / YSO
& stellar / star--disk corona
& X-ray flare; accretion burst
& protostellar jet / outflow
& episodic accretion and outflow recurrence; no generic solar-like cycle
& reported in some data
& star--disk dynamo + episodic accretion + reconnection \\
Sun
& solar corona
& solar flare
& CME
& 11-year activity cycle (22-year cycle)
& clear
& $\alpha\Omega$ dynamo + flux transport + reconnection \\
Active stars \cite{Maehara2012,Maehara2015,Doyle2018MNRAS}
& stellar corona
& stellar flare / superflares
& stellar CME candidates
& magnetic activity cycles in many stars
& reported some 
& stellar dynamo \\
Earth
& magnetosphere
& substorm
& plasmoid tail jet / bursty bulk flow
& secular variation, excursions, and reversals; no stable periodic cycle
& present in geomagnetic indices
& internal geodynamo + solar-wind coupling \\
Jupiter and giant planets \cite{Manners2020JGR,Gershman2019GRL}
& magnetosphere
& auroral burst / injection event
& magnetotail jet / plasmoid ejection
& rotation-modulated and secular variability; no solar-like cycle
& broadband / ULF fluctuations reported
& internal dynamo + rapid rotation + external plasma loading / solar-wind forcing \\
Laboratory dynamo (VKS) \cite{Monchaux2007PRL,Berhanu2007EPL,Giesecke2010PRL,Herault2016EPJ}
& nonequilibrium conducting-fluid volume
& magnetic burst / excursion
& impulsive global field reorganization
& irregular polarity reversals, excursions, and stationary / oscillatory dynamo regimes
& reported in magnetic induction time series
& turbulent liquid-sodium dynamo + mode competition + boundary-condition effects (especially soft-iron impellers) \\
\hline
\end{tabular}%
}
\end{sidewaystable*}


\end{enumerate}

\section{Conclusions}
We have proposed the Synchronized Spin Model as a mesoscopic nonequilibrium framework for
black-hole accretion systems. The key claim is that a rotating magnetized accretion flow can be
coarse-grained into interacting magnetic domains in a Taylor-column-like picture whose synchronization, partial synchronization, excursion, and reversal govern both the timing statistics and the magnetic topology of the
flow.

The model yields three main conclusions. 
First, partial synchronization and amplitude modulation provide a natural route to pink-noise continua, 
rms--flux/Taylor-like
scaling, and approximately log-normal variability of the demodulated envelope. 
Second, a small
set of collective variables,
\begin{equation}
(M, G, f_{\rm open}, C_{\rm ax}),
\end{equation}
organizes the main morphological channels: closed topology favors the corona, partial coherent
reversal favors flares, open but weakly axial topology favors winds, and ordered axial open topology favors jets. 
Third, the hard/soft/intermediate cycle can be read as motion through magnetic
state space, with the hard branch retaining stronger macro-order, the jet-line region marking
rapid restructuring, and the soft branch suppressing steady jets because its global dipole-like
order is weaker.

The broader significance of SSM is to add a compact state-theory layer between first-principles
simulations and phenomenological observables. Its purpose is not to replace GRMHD or launching theory, but to connect variability statistics, magnetic topology, and outflow morphology
within one framework.

In this revised statistical reading, the continuous variability of black-hole systems is more naturally characterized by multiplicative or approximately log-normal fluctuations than by a generic
power-law flux distribution. By contrast, power-law-like event statistics remain a useful effective language only for suitably defined event catalogs or reconnection hierarchies, and are
most directly established in solar and stellar flare systems. This distinction clarifies how the
SSM can accommodate both rms--flux/log-normal phenomenology and broader multiscale reconnection activity without conflating continuous light-curve statistics with event-by-event energy distributions.
In this sense, the SSM provides a falsifiable mesoscopic framework for black-hole accretion systems, whose main value lies in connecting timing statistics, magnetic topology, and outflow morphology within a single reduced-state language.

\section*{Acknowledgments}
We are grateful to Kazuo Makishima, Ryoji Matsumoto, Tatehiro Mihara, Shin-ya Nitta, Motoko Serino, Kazunari Shibata, Masaaki Takahashi, Miho Tan, and many members of Lunch Meeting for fruitful discussions.

\printbibliography

\end{document}